\documentclass[epj]{webofc}
\usepackage[varg]{txfonts}   
\usepackage{graphicx,color} 
\usepackage{bm}       
\usepackage{amsmath}  
\usepackage{amssymb}
\woctitle{21st International Conference on Few-Body Problems in Physics}
\begin{document}
\title{Weakly bound states with spin-isospin symmetry}
\author{A. Kievsky\inst{1}\fnsep\thanks{\email{kievsky@pi.infn.it}} \and
        M. Gattobigio\inst{2}\fnsep\thanks{\email{mario.gattobigio@inln.cnrs.fr}}
}

\institute{Istituto Nazionale di Fisica Nucleare, Largo Pontecorvo 3, 56127 Pisa, Italy
\and
Universit\'e de Nice-Sophia Antipolis, Institut Non-Lin\'eaire de
Nice,  CNRS, 1361 route des Lucioles, 06560 Valbonne, France
          }

\abstract{
We discuss weakly bound states of a few-fermion system having spin-isospin symmetry.
This corresponds to the nuclear physics case in which the singlet, $a_0$, and triplet,
$a_1$, $n-p$ scattering lengths are large with respect to the range of the nuclear interaction.
The ratio of the two is about $a_0/a_1\approx-4.31$. This value defines a plane in which $a_0$ and
$a_1$ can be varied up to the unitary limit, $1/a_0=0$ and $1/a_1=0$, maintaining its ratio fixed.
Using a spin dependant potential model we estimate the three-nucleon binding energy along
that plane. This analysis can be considered an extension of the
Efimov plot for three bosons to the case of three $1/2$-spin-isospin fermions.
}
\maketitle
\section{Introduction}
\label{intro}

In a series of papers~\cite{efimov1,efimov2}
V. Efimov has shown that three identical bosons interacting through a two-body short-range potential
have a resonant spectrum at the unitary limit, $1/a=0$, with $a$ the two-body scattering
length. In this limit a geometrical series of bound states appears whose energies accumulate to zero.
The ratio between the energies of two consecutive states is constant and, remarkably it results
independent of the particular form of the interaction. This behavior has been denoted
Efimov effect and its characteristic of being universal has given to this effect a
particular relevance. In fact in the last two decades an enormous amount of work has
been dedicated to study this effect in
different fields as molecular, atomic, nuclear and particle physics.

The spectrum of the three-boson system close to the unitary limit is described by the Efimov
equations (or Efimov radial law) that can be put in the following form

\begin{equation}
  \begin{gathered}
    E_3^n/(\hbar^2/m a^2) = \tan^2\xi \\
    \kappa_*a = \text{e}^{(n-n^*)\pi/s_0} 
    \frac{\text{e}^{-\Delta(\xi)/2s_0}}{\cos\xi}\,,
  \end{gathered}
  \label{eq:energyzr}
\end{equation}
where $\Delta(\xi)$ is a one parameter universal function depending on the ratio
of the three- and two-body binding energies. A parametrization of this
function in the range $-\pi < \xi < -\pi/4$ can be
found for example in Ref.~\cite{report}.
The quantity $s_0\approx 1.00624$ is a universal number and
$\kappa_*$ defines the energy
$\hbar^2 \kappa_*^2/m$ of the level $n=n^*$ at the unitary limit. 
Furthermore, knowing the
value of $\kappa_*$ the complete spectrum is determined as a function of $a$.
It should be noticed however that the above equations have been derived
in the zero-range limit. In this limit the two-body energy results to be
$E_2=\hbar^2/ma^2$ and, in addition, all the two-body low energy scattering
observables can be written in terms of $a$, making this quantity a control
parameter (see for instance Ref.~\cite{report} and references therein).
When a two-body potential with finite range is considered, the Efimov
radial law can be modified as it was discussed by the authors in
Refs.~\cite{gatto1,kievsky1}. In the following we extend the discussion
to the case in which the particles are $1/2$-spin-isospin fermions. 

\section{Efimov plot with $1/2$-spin-isospin fermions}
\label{sec-1}

The zero-range energy spectrum given by Eq.(\ref{eq:energyzr}) is unbounded
from below as has been shown first by Thomas~\cite{thomas}. Finite-range 
potentials cure this pathology modifying the energy spectrum.
For values of $a$ much bigger than the range of the interaction 
Eq.(\ref{eq:energyzr}) is still valid for the highest levels, however
the ground and first excited levels show finite-range effects. In order
to take into account these corrections the authors have proposed a
modification to the Efimov radial low~\cite{gatto1,kievsky1},

\begin{equation}
  \begin{gathered}
    E_3^n/E_2 = \tan^2\xi \\
    \kappa^3_na_B + \Gamma_n^3 =
    \frac{\text{e}^{-\Delta(\xi)/2s_0}}{\cos\xi} \,.
  \end{gathered}
  \label{eq:energyfr}
\end{equation}
Here there are two corrections, one coming from the two-body system
and it can be taken into account by substituting $a$ with $a_B$, defined by
$E_2=\hbar^2/m a_B^2$, with $E_2$ the two-body binding energy if $a>0$, or the two-body
virtual-state energy in the opposite case, $a<0$.
The second correction is the introduction of a finite-range parameter,
$\Gamma_n^3$, depending on the energy level, which produces a 
shift in the product $\kappa_*a_B$. Recently it has been shown that the
value of the shift is almost the same for very different potentials in range
and scale as the LM2M2 interaction of Aziz~\cite{aziz}, a gaussian potential
reproducing the values of $E_2$ and $a$ given by that interaction and
a combination of yukawians, the MT-III interaction~\cite{tjon}, describing the $s$-wave 
triplet $n-p$ state. Interestingly the value of the shift for the ground state
in the three cases results to be almost the same, $\Gamma_0^3\approx 0.8$~\cite{kievsky2015}.

In the following we would like to discuss the case of fermions having $1/2$ isospin
symmetry in which the interaction is different in the singlet and triple state as 
for the $n-p$ system. In this case the experimental data are $a_0=-23.740\pm0.020\;$fm
and $a_1=5.424\pm0.003\;$fm. To describe these data we use a gaussian potential
\begin{equation}
V(r)=V_S {\rm e}^{-r^2/r_S^2}\, ,
\end{equation}
with different strength in the singlet ($S=0$) and triplet ($S=1$) and having the
same range (in the following we consider $r_0=r_1=1.65\;$ fm). The values
$V_0=-37.90$ MeV and $V_1=-60.575$ MeV approximate the experimental data and,
for $S=1$, the deuteron binding energy is approximated too. Varying the strengths
$V_S$ it is possible to explore the three-nucleon binding energy as the values
of $a_S$ move toward the unitary limit. Among different possibilities we fix
the ratio $a_0/a_1=-4.31$ to its experimental value and calculate the energy of
the ground and first excited states. In Fig.1 we show the different planes
(the three-body energy defines the third axis not shown in the figure)
determined by the values of $a_0$ and $a_1$. When the two scattering lengths are
equal, $a_0/a_1=1$, the corresponding plane is the boson plane for three equal
particles very well studied (see for instance Ref.~\cite{gatto1}). In the figure
this is given by the (red) solid line. The (blue) solid line corresponds to what
we call the nuclear plane since it contains the physical nuclear point, the 
continuation to the fourth quadrant given by the (blue) dashed line does not
give new information since the exchange of the singlet and triplet potentials
gives the same binding energy. The two axes correspond to planes in which one
of the scattering lengths is on the unitary limit. The energy values of $E_3^n$
are different in the different planes however they collapse to the boson spectrum
when the two scattering lengths are at the unitary limit. Using the gaussian
potential given above the ground state energy in this limit
results $E_3^0\approx 3.6\;$ MeV.

\begin{figure}
\centering
\includegraphics[width=8cm,clip]{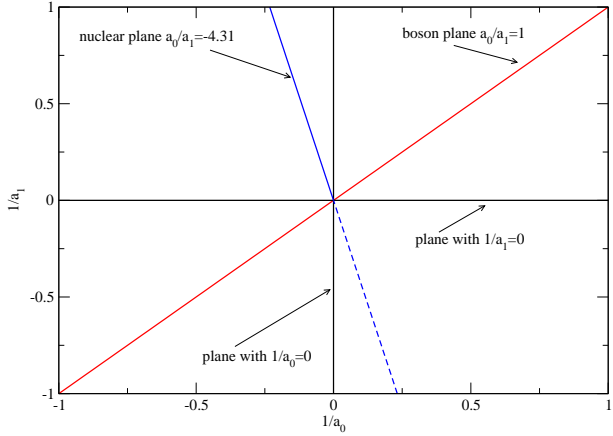}
\caption{Different planes as function of the inverse of $a_0$ and $a_1$}
\label{fig-1}       
\end{figure}

The main results of the present study is given in Fig.2 where the momentum binding energies,
defined as $(\hbar^2/m)K^2_3=E_3$, 
for the ground and excited states are shown as a function of $a_1$ (in units
of $r_*$, the value of the effective range at the unitary limit) maintaining the
ratio $a_0/a_1$ fixed around $-4.31$. The three-fermion momentum binding energies are given
by the (blue) solid line for the ground state and by the (blue) dashed line
for the first excited state. For comparison the momentum energy values in the case of three
bosons are given too by the (black) solid and dashed lines. The (red) solid line is
the two-body momentum biding energy. $K_*$ is the momentum at the unitary limit and both, the
fermion and boson values, are the same. The excited state in this limit has a
value of $K_*/22.9$ showing a slight finite-range correction (the zero-range theory predicts
the value $K_*/22.7$). The (negative) values of the triplet scattering lengths, 
$a^0_-$ and $a^1_-$, at which the three-boson ground and excited states disappear into the 
three-nucleon continuum are shown in the figure. Their ratio of about $17.6$ is well below
the prediction of $22.7$ given by the zero-range theory. More interesting is the fact that
the excited state of the three-fermion system disappears into the $n-d$ continuum before
matching the nuclear physical point (at $a^1_*\approx 20\;$fm). This is not the case for the
three-boson system, the excited state remains bound and always below $E_2$. This fact
has been studied before in the case of a system of three helium atoms. The present analysis
shows that in the case of three nucleons the difference in the single and triple potential 
strengths is such that the excited state disappear into the $n-d$ continuum before reaching
the physical point. Evidence of the presence of this state embedded in the $n-d$ continuum has 
been given in the literature~\cite{fonseca,tobias} and the curvature of the $n-d$ effective 
range function close to the threshold is one of this~\cite{friar,kievsky}. Finally, at the
physical value, the two- and three-body energies are explicitly displayed. 
As it is well known, in order
to reproduce the triton binding energy a three-body force has to be included.

The spectrum of the three-fermion system can be still described by Eq.(\ref{eq:energyfr})
with a value of the shift depending on the ratio $a_0/a_1$. In the present analysis we
found for the ground state level $\Gamma^0_3\approx -0.2$, a negative value. If we compare 
this value to the value of about $0.8$ obtained in the case of three bosons we can conclude
that there is a particular value of the ratio at which the shift $\Gamma^0_3\approx 0$.
In this case Eq.(\ref{eq:energyfr}) will be equivalent to the zero-range equation. However
this would valid for the ground state level, the shift depends on the level and, 
as we have shown, the ratio with the excited state at the unitary limit is $22.9$ (not $22.7$) 
for a gaussian potential independently of the shift value.

\begin{figure}
\centering
\includegraphics[width=8cm,clip]{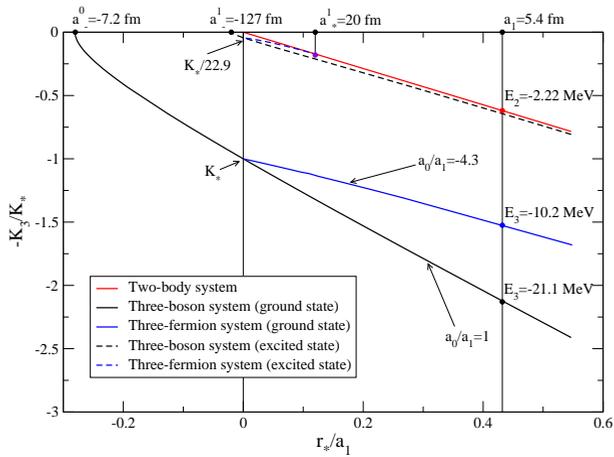}
\caption{The momentum energy as a function of the inverse of $a_1$}
\label{fig-2}       
\end{figure}

\section{Conclusions and Outlook}

The present analysis focussed on the particularities of the Efimov plot in the
case of $1/2$-spin-isospin fermions. We have selected a particular way of
constructing the Efimov plot in which the ratio of the singlet and triplet
scattering lengths has been kept fixed. In this way the momentum binding
energy and $a_1$ define a plane similar to the boson case, in particular the
values at the unitary limit coincide. The present work has to be consider a first
step in the study of the Efimov plot for three nucleons. At present the role
of the three-body force and the predictions on systems with $A>3$ are underway.

\end{document}